\documentclass[aps,prb,twocolumn,longbibliography,superscriptaddress,floatfix,10pt]{revtex4-2}

\usepackage[utf8]{inputenc}
\usepackage{amssymb,amsfonts,amsmath,dsfont}
\usepackage{graphicx}
\usepackage{textcomp}
\usepackage{gensymb}
\usepackage{pifont}
\usepackage{bm}
\usepackage{color}
\usepackage{accents}
\usepackage{mathtools}
\usepackage{nicefrac}
\usepackage{url}
\usepackage{multirow}
\usepackage{hyperref}
\usepackage[capitalize]{cleveref}
\usepackage{tabularx}
\usepackage{upgreek}
\usepackage{chemformula}
\hypersetup{colorlinks=true,citecolor=blue,urlcolor=blue,linkcolor=blue}

\usepackage{mathrsfs}

\newcommand{\bs}[1]{\boldsymbol{#1}}

\makeatletter

\begin{document}

\title{Theory of magnon hydrodynamics in collinear antiferromagnets}
\author{Vivianne Olguín-Arias}
\affiliation{Departamento de Física, Facultad de Ciencias, Universidad de Chile, Chile}

\author{Alireza Qaiumzadeh}
\affiliation{Center for Quantum Spintronics, Department of Physics, Norwegian University of Science and Technology, NO-7491 Trondheim, Norway}

\author{Roberto E. Troncoso}
\affiliation{Departamento de Física, Facultad de Ciencias, Universidad de Tarapacá, Casilla 7-D, Arica, Chile}

\begin{abstract}
We develop a theoretical framework for the transport of spin angular momentum and linear momentum carried by magnons in electrically insulating collinear antiferromagnets (AFs). Focusing on both transverse and longitudinal geometries, we model magnons as a viscous fluid and examine the hydrodynamic regime that arises when the momentum-conserving magnon–magnon scattering length is shorter than the momentum–relaxation length associated with momentum-nonconserving processes.
The hydrodynamic regime in these two geometries is marked by distinct flow patterns: vortex structures in the transverse setup and Poiseuille flow in the longitudinal setup.
We show that scattering between magnon modes, carrying opposite spin angular momentum, induces drag-like effects, significantly altering magnon spin current propagation in AFs. Furthermore, we uncover mode-dependent transport regimes at specific distances by tuning the magnetic field, which enables selective control of viscous and diffusive transport in individual magnon modes.
Our results position AFs as a promising platform for realizing magnon-fluid dynamics and studying collective spin transport phenomena.
\end{abstract}

\maketitle
\section{Introduction}
The fluid dynamics of liquids is governed by hydrodynamic equations; the conservation laws of mass, momentum, and energy \cite{Forster2018,Chapman}. This is a fundamental discipline that spans across engineering, theoretical, and applied sciences. The hydrodynamic regime is established when particle collisions are sufficiently strong and the mean free path for momentum-conserving collisions is much smaller than the system size, ensuring local thermodynamical equilibrium \cite{Polini2020}. These principles extend to systems of electrons, where at low temperatures or high densities, macroscopic properties such as electrical conductivity and heat flow behave similarly to classical fluids \cite{Narozhny2022,Berdyugin2019}. Electron hydrodynamics occurs when electron collisions dominate over non-momentum-conserving scattering events, e.g., with impurities, phonons, or lattice defects \cite{Narozhny2022}, the collective dynamics governed by the viscosity. 
Signatures of electron fluids have recently been observed in a wide variety of materials like mono-\cite{Crossno2016,Bandurin2016} and bilayer graphene \cite{Nam2017,Bandurin2018}, (Al,Ga)As \cite{Jong1995,Buhmann2002,Ginzburg2021}, WTe${}_2$ \cite{Vool2021,AharonSteinberg2022}, WP${}_2$ \cite{Gooth2018}, and PdCoO${}_2$ \cite{Moll2016}, exhibiting phenomena like the Gurzhi effect \cite{Gurzhi1968}, whirlpools \cite{AharonSteinberg2022,Palm2024,Bandurin2016}, and the formation of Poiseuille’s flow \cite{Sulpizio2019,Jenkins2022}. Recently, the hydrodynamic paradigm has extended to various bosonic systems, such as phonons \cite{Cepellotti2015,Ghosh2022} and magnons \cite{UlloaPRL2019,Rodriguez-Nieva2022,xue2024}.

Magnons, bosonic quasiparticles representing collective spin fluctuations in magnetically ordered systems, are electrically neutral and carry both spin angular momentum and linear momentum. These collective spin excitations are at the heart of magnonics \cite{Pirro2021}, which encompass different uses of the spin degrees of freedom heading to the development of quantum computation and communication technologies \cite{Magnonics2013}. Thus, understanding their transport behavior—whether diffusive, ballistic, or hydrodynamic—is crucial for designing magnon-based devices \cite{Flebus2023}. In the diffusive regime, magnon transport is governed by random scattering events (e.g., phonon, dislocation, impurity, and boundary scattering) that rapidly destroy momentum but allow for a net flux of spin or heat driven by gradients (e.g., temperature, chemical potential, or spin accumulation) \cite{Cornelissen2016,Troncoso2020}. This regime applies when momentum-relaxing scattering processes dominate over momentum-conserving magnon-magnon interactions. As a result of complex processes of interactions, the transport of magnon currents is described by drift-diffusion equations, where the spin diffusion length determines how far spin signals might propagate. Within this framework, the spin Seebeck effect \cite{Xiao2010,Adachi2013,Rezende2014} and nonlocal magnon transport experiments \cite{Cornelissen2015,Lebrun2018} are explained.
\begin{figure}[tbh]
\includegraphics[width=85mm]{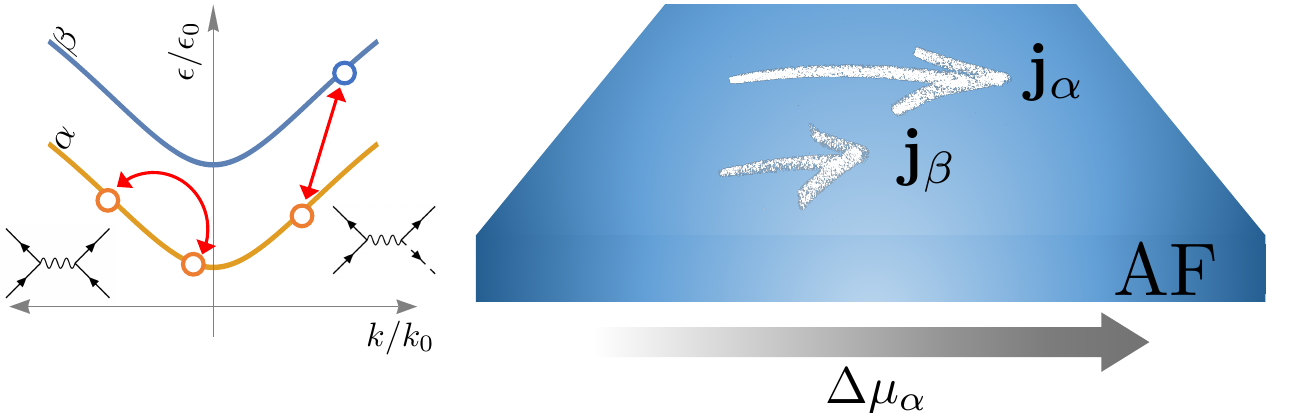}
\caption{Schematic description for the fluid-like dynamics of magnon currents in a collinear AF. Magnonic bands are splitted because of an applied  Zeeman field. Due to interconversion of chiral antiferromagnetic magnons, and viscous effects, linear momentum is transferred among the pair of chiral $\alpha-$ and $\beta-$magnon subsystem. Under an external driving, e.g., induced by a spin Hall effect (SHE), a magnon spin current ${\bs j}_{\alpha}$ is accompanied by the dragged magnon spin current ${\bs j}_{\beta}$.}
	\label{fig:setup}
\end{figure} 

When the momentum-conserving magnon-magnon scattering process dominates over momentum-relaxing processes, magnons may enter a hydrodynamic regime. In this form of transport, magnonic excitations mimic viscous fluid behaviors, being its well-defined local average velocity described by the Navier-Stokes equations, has been scarcely scrutinized. The relevant length scale is the momentum-relaxation length and hydrodynamics becomes applicable when the system size is much larger than the momentum-relaxation length, so that collective flow can develop. Collective hydrodynamic and viscous transport has been studied in ferromagnetic insulators \cite{Sano2023,UlloaPRL2019,Li2022}, showing a clear signal of the presence of whirlpools, a hallmark of hydrodynamic flow, in nonlocal transport measurements. Poiseuille flow has been proposed to explain a magnon mean-free path in observations on Cu${}_2$OSeO${}_3$ \cite{Prasai2017}. In yttrium-iron garnet (YIG) films, by the spin transfer effect near the Curie temperature, a two-fluid behavior comprising high- and low-energy magnons \cite{KohnoPRB2023-1,KohnoPRB2023-2} has been observed. Recently, it has been proposed that local magnetometers, e.g., SQUIDs and nitrogen-vacancy centers, detect hydrodynamic signatures and viscosity-induced structure in the stray magnetic field \cite{xue2024,Son2025}.

In this paper, we investigate fluid-like dynamics of magnons in uniaxial antiferromagnetic insulators. AFs meet unique properties such as stray-field immunity and ultra-fast response dynamics, relevant for spintronics \cite{Jungwirth2016,Baltz2018rmp}. Unlike in ferromagnets, magnons in uniaxial easy-axis AFs exist in two chiral flavors carrying opposite spin angular momentum $\pm\hbar$. Although several magnon transport regimes in AFs, including magnon diffusion \cite{Rezende_2018,PhysRevB.107.184404,Troncoso2020,PhysRevB.110.L140408}, magnon superfluidity \cite{PhysRevB.90.094408, PhysRevLett.118.137201}, and magnon Bose–Einstein condensation \cite{PhysRevB.111.184402}, have been explored theoretically in recent years, a unified and comprehensive theoretical framework for the magnon hydrodynamics regime in these systems is still lacking.
The underlying hydrodynamic phenomena in the two-band system rest upon conservative scattering processes between magnons that transfer spin- and linear-momentum, causing drag and viscous effects. Magnonic drag is due to interband scatterings that originate from the exchange energy and magnetocrystalline anisotropy, while viscous effects are induced by momentum and spin-conserving intraband scattering events from exchange energy. These effects take place simultaneously with mechanisms that lead to the decay of the magnon number and the diffusion of magnon currents. Therefore, we determine the hydrodynamic equations from semiclassical Boltzmann theory, with a focus on the crossover between diffusive and hydrodynamic regimes. Finally, the fluid response is solved in a nonlocal spin transport geometry and in a normal metal-AF-normal metal heterostructure.

The paper layout is as follows. In Sec. II, we present the spin Hamiltonian model and the corresponding theory for interacting magnons. In Sec. III, we formulate the semiclassical Boltzmann theory for AF magnons, to determine in Sec. IV the hydrodynamic equations for magnon(spin)-density and velocity, i.e., the diffusion and Navier-Stokes equations. These equations are solved in the setups depicted at Fig. \ref{fig: magnon-transport}. Finally, in Sec. V, we provide the conclusions and discuss the implications raised in our work.

\section{Spin model and magnons}
We consider a three-dimensional (3D) collinear uniaxial antiferromagnetic insulator composed of localized spins arranged on a cubic lattice. The system is described by a nearest-neighbor spin Hamiltonian with two antiparallel magnetic sublattices,
\begin{align}\label{eq:spinH}
H_{\text{AF}}=J\sum_{\left\langle {i}{j}\right\rangle }{\mathbf{S}}_{{i}}\cdot{\mathbf{S}}_{{j}}-\frac{\kappa_z}{2S}\sum_{{i}}{{S}}_{{i}z}^{2}-h\sum_{{i}}{S}_{{i}z},
\end{align}
with ${\mathbf{S}}_{i}$ the spin vector operator at site ${i}$, $J>0$ the antiferromagnetic Heisenberg exchange coupling, $\kappa_z>0$ the strength of the uniaxial easy-axis magnetic anisotropy, and the Zeeman interaction due to an external magnetic field $h$ applied along the $z$-axis.

To describe low-energy magnons, we derive the nonlinear Hamiltonian governing magnon-magnon interactions \cite{Rezende2019, Shiranzaei_2022}. Using the standard Holstein–Primakoff (HP) transformation \cite{HolsteinPR1940}, spin fluctuations on the two antiparallel ${\cal A}$ and ${\cal B}$ sublattices are respectively mapped onto bosonic fields $a_i$ and $b_i$ that quantify deviations from the classical Néel ground state,
\begin{align}
{S}^{+}_{i,\cal A}=\sqrt{2S-a^{\dagger}_{i}a_{i}}\,a_{i},\quad {S}^{+}_{{i},\cal B}=b^{\dagger}_{i}\,\sqrt{2S-b^{\dagger}_{i}b_{i}},
\end{align}
moreover, ${S}^{z}_{{i},\cal A}=S-a^{\dagger}_{i}a_{i}$ and  ${S}^{z}_{{i},\cal B}=-S+b^{\dagger}_{i}b_{i}$, where the rising and lowering operators satisfy ${{S}}^{-}_{i,\cal A}=({{S}}^{+}_{i,\cal A})^{\dagger}$, ${S}^{-}_{{i},\cal B}=(\hat{S}^{+}_{{i},\cal B})^{\dagger}$, and $S$ is the spin quantum number. Applying a canonical Bogoliubov and Fourier transformation yields new bosonic magnon operators that diagonalize the quadratic part of the Hamiltonian. The resulting interacting Hamiltonian takes the form $H_{\text{AF}} = H_0 + H_I$, where the diagonalized quadratic Hamiltonian in the $\{\alpha, \beta\}$ bosonic magnon basis reads,
\begin{align}\label{eq:freemagnonH}
H_{0}=\sum_{\xi\in\{\alpha,\beta\}} \sum_{{\bs k}}\epsilon_{\xi,\bs k}{\xi}^{\dagger}_{\bs k}{\xi}_{\bs k},
\end{align}
where each magnon eigenmode, $\alpha$ and $\beta$, carries opposite ($\pm\hbar$) spin-angular momentum with eigenenergies $\epsilon_{\xi,\bs k}=\pm h+\sqrt{(6JS)^2(1-\gamma^2_{\bs k})+H^2_c}$, with $\gamma_{\bs q}$ the structure factor and $H_c=\kappa_z^2+12\kappa_z JS$ being the critical spin-flop field \cite{Gurevich2020}. The interacting part of the Hamiltonian, obtained from the fourth-order expansion of $H_{\text{AF}}$, describes collisions that represent the lowest order non-linear processes that take place in the magnetic dynamics. We split $H_I$ into two contributions, the intraband scattering processes ($V_1$) which conserve the number of magnons in each eigenstate,
\begin{align}\label{eq:interactingV1}
V_1=\sum_{\xi\in\{\alpha,\beta\}}\sum_{{\bs q}{\bs k}{\bs k}'}{\textsc v}^{(\xi)}_{{\bs q}{\bs k}{\bs k}'}{\xi}^{\dagger}_{\bs k+\bs q}{\xi}^{\dagger}_{\bs k'-\bs q}{\xi}_{\bs k}{\xi}_{\bs k'},
\end{align}
and the interband collisions ($V_2$) which do not conserve the number of magnons in individual eigenmodes,
\begin{align}\label{eq:interactingV2}
\nonumber V_2=\sum_{{\bs q}{\bs k}{\bs k}'}&\left[u_{{\bs q}{\bs k}{\bs k}'}\alpha^{\dagger}_{{\bs k}+{\bs q}}\beta_{{\bs q}-{\bs k}'}\alpha_{{\bs k}}\alpha_{{\bs k}'}+v_{{\bs q}{\bs k}{\bs k}'}\alpha^{\dagger}_{{\bs k}+{\bs q}}\beta_{{\bs k}'+{\bs q}}\alpha_{{\bs k}}\beta^{\dagger}_{{\bs k}'}\right.\\
&\left.+w_{{\bs q}{\bs k}{\bs k}'}\beta_{{\bs k}'-{\bs q}}\beta_{{\bs q}-{\bs k}}\alpha_{{\bs k}}\beta^{\dagger}_{{\bs k}'} +\text{h.c.}\right],  
\end{align}
written in the corresponding eigenbasis. The scattering amplitude in the intraband collisions between magnons are ${\textsc v}^{(\xi)}_{{\bs q}{\bs k}{\bs k}'}$, while for interband interactions are $u_{{\bs q}{\bs k}{\bs k}'}$, $v_{{\bs q}{\bs k}{\bs k}'}$, and $w_{{\bs q}{\bs k}{\bs k}'}$, given at Appendix \ref{app:scattamp}. Spin-angular momentum is also conserved in these scattering processes, which is manifestly due to $U(1)$ axial symmetry of the spin Hamiltonian. Higher order magnon collisions become relevant at high temperatures or at the verge of magnetic phase transitions \cite{Rezende2019}, which is not the case here.

\section{Semiclassical Boltzmann theory}
The system of antiferromagnetic magnons is assumed in a hydrodynamic regime, which is guaranteed when momentum-conserving and spin-conserving magnon-magnon collisions are dominant over momentum-nonconserving scattering processes, e.g., with phonons, disorder, or impurities \cite{Narozhny2022}. The existence of a microscopic relaxation time $\tau$ features the approach to local equilibrium, parametrized by a magnon chemical potential $\mu_{\xi}({\bs x})$ and an effective temperature $T_{\xi}({\bs x})$, for each magnon $\xi=\{\alpha, \beta\}$ mode. Thus, at the hydrodynamic regime, the Boltzmann equation for the nonequilibrium magnons distribution function $f_{\xi}\left({\bs x},{\bs k};t\right)$ satisfies,
\begin{align}
\frac{\partial f_{\xi}}{\partial t}+{\bf v}_{\xi\bs k}\cdot\frac{\partial f_{\xi}}{\partial {\bs x}}+{\bs F}\cdot\frac{\partial f_{\xi}}{\partial {\bs k}}&\label{eq:BEalphabeta}={\cal C}^{\xi\xi}_{\bs k}+{\cal C}^{\xi\xi'}_{\bs k},
\end{align}
with ${\bf v}_{\xi\bs k}={\partial_{\bs k}\epsilon_{\xi,\bs k}}$ being the magnon group velocity, and ${\bs F}$ an external driving force. The net flux of magnons into the state ${\bs k}$ is given by the intraband and interband collision integrals, ${\cal C}^{\xi\xi}_{\bs k}$ and ${\cal C}^{\xi\xi'}_{\bs k}$, respectively, which are detailed as follows. For a magnon mode $\xi$, the intraband collision integral expands as ${\cal C}^{\xi\xi}_{\bs k}={\cal C}^{\xi,\text{el}}_{\bs k}+{\cal C}^{\xi,\text{mr}}_{\bs k}+{\cal C}^{\xi,\text{mp}}_{\bs k}+{\cal C}^{\xi,\text{mm}}_{\bs k}$, containing spin-nonconserving and conserving processes. Spin-nonconserving processes comprise elastic magnon-impurity collisions (${\cal C}^{\xi,\text{el}}_{\bs k}$) and magnon relaxation (${\cal C}^{\xi,\text{mr}}_{\bs k}$) due to Gilbert damping, while momentum- and spin-conserving processes contribute in the form of intraband magnon-magnon collisions (${\cal C}^{\xi,\text{mm}}_{\bs k}$) and magnon-phonon processes due to modulation of exchange (${\cal C}^{\xi,\text{mp}}_{\bs k}$). On the other hand, the interband collision integral ${\cal C}^{\xi\xi'}_{\bs k}$, evaluated within the framework of Fermi’s golden rule, is given by,
\begin{widetext}
\begin{align}
{\cal C}^{\alpha\beta}_{\bs k}=&\frac{2\pi}{\hbar}\sum_{{\bs q}{\bs k}'}\nonumber\left\{u^2_{{\bs k}{\bs q}{\bs k}'}\delta(\epsilon_{\alpha{\bs k}}+\epsilon_{\beta,{\bs q}-{\bs k}'}+\epsilon_{\alpha{\bs k}'}-\epsilon_{\alpha,{\bs k}+{\bs q}})\left[\left(1+f^{\alpha}_{{\bs k}}\right)(1+f^{\beta}_{{\bs q}-{\bs k}'})\left(1+f^{\alpha}_{{\bs k}'}\right)f^{\alpha}_{{\bs k}+{\bs q}}-f^{\alpha}_{{\bs k}}f^{\beta}_{{\bs q}-{\bs k}'}f^{\alpha}_{{\bs k}'}\left(1+f^{\alpha}_{{\bs q}+{\bs k}}\right)\right]\right.\\
&\label{eq:interband-collision1}\left.+w^2_{{\bs k}{\bs q}{\bs k}'}\delta(\epsilon_{\alpha{\bs k}}+\epsilon_{\beta,{\bs k}'-{\bs q}}+\epsilon_{\beta,{\bs q}-{\bs k}}-\epsilon_{\beta{\bs k}'})\left[\left(1+f^{\alpha}_{{\bs k}}\right)(1+f^{\beta}_{{\bs k}'-{\bs q}})(1+f^{\beta}_{{\bs q}-{\bs k}})f^{\beta}_{{\bs k}'}-f^{\alpha}_{{\bs k}}f^{\beta}_{{\bs k}'-{\bs q}}f^{\beta}_{{\bs q}-{\bs k}}(1+f^{\beta}_{{\bs k}'})\right]\right\}.
\end{align}
\end{widetext}
It describes the exchange of quasiparticles between the $\alpha$ and $\beta$ magnon subsystems, arising from microscopic processes governed by the interacting Hamiltonian in Eq. (\ref{eq:interactingV2}). These processes involve the creation or annihilation of magnon pairs, with each event mediated by the interaction amplitudes $u_{{\bs q}{\bs k}{\bs k}'}$ and $w_{{\bs q}{\bs k}{\bs k}'}$, detailed at the appendix \ref{app:scattamp}. The chemical potential of each magnon subsystem is modified, resulting in a transfer of linear momentum between the $\alpha$ and $\beta$ magnon modes. The collision integral ${\cal C}^{\xi\xi'}_{\bs k}$ is not symmetric, in general, under a swap of $\xi$ and $\xi'$ labels, if both magnon modes are not reciprocal. It can be the case when a magnetic field is applied, where the degeneracy of $\alpha$ and $\beta$ magnons is removed. 

We now determine the magnon spin currents in the inviscid regime from the linearized Boltzmann equation, Eq. (\ref{eq:BEalphabeta}). We begin by assuming that momentum-nonconserving scattering processes dominate over momentum-conserving scattering and magnon–magnon collisions that conserve the total magnon number. We assume that strong inelastic spin-preserving processes fix the magnon effective temperatures to that of the local phonon bath temperature; therefore, only the magnon chemical potentials $\mu_\alpha({\bs x})$ and $\mu_\beta({\bs x})$ will then be determined. Within the relaxation time approximation, the collision integrals ${\cal C}^{\xi\xi}_{\bs k}$ and ${\cal C}^{\xi\xi'}_{\bs k}$ govern, respectively, the scattering times $\tau_{\xi\xi}$ and $\tau_{\xi\xi'}$, thus quantifying the relaxation towards the thermodynamical equilibrium \cite{Cornelissen2016,Troncoso2020}. These are defined through the out-collision rates  as
\begin{align}
\frac{1}{\tau_{\xi\xi'{\bs k}}}=\frac{{\cal C}^{\xi\xi'}_{{\bs k},{\rm out}}}{f_{\xi}({\bs k},t)},
\end{align}
here, $\tau_{\xi\xi'}$ denotes the scattering time between magnon modes $\xi$ and $\xi'$, where $\xi \ne \xi'$ corresponds to the interconversion between $\alpha$- and $\beta$-magnons, and $\xi = \xi'$ describes intramode relaxation.

Next, we linearize the out-of-equilibrium distribution of chiral magnons. The distribution for the $\xi$-magnon mode is represented by $f_{\xi}\left({\bs x},{\bs k},t\right)=f^0_{\xi}\left({\bs x},{\bs k}\right)+\delta f_{\xi}\left({\bs x},{\bs k},t\right)$, where $\delta f_{\xi}$ describes deviations from the Bose-Einstein distribution $f^0_{\xi}({\bs x},{\bs k})=\left(e^{\left({\epsilon_{\xi\bs k}-{\mu}_{\xi}({\bs x})}\right)/{k_BT_{\xi}({\bs x})}}-1\right)^{-1}$, which is parametrized by the local chemical potential ${\mu}_{\xi}(\bs x)$ and the effective temperature $T_{\xi}(\bs x)$. Within this approximation, the collision integrals are linearized in terms of the distributions $\delta f_{\alpha}$ and $\delta f_{\beta}$. For the intraband collisions, we get ${\cal C}^{\xi\xi}_{\bs k}= -\sum_{p \in \Omega} \delta{f_{\xi}}/{\tau_{p}}$, where $\Omega$ contains all scattering processes introduced above \cite{Cornelissen2016}, while for the interband collisions ${\cal C}^{\alpha\beta}_{\bs k}= -\delta f_\alpha/{\tau_{\alpha\beta}} - \delta{f_\beta}/{\bar{\tau}_{\beta\alpha}}$ and ${\cal C}^{\beta\alpha}_{\bs k}=-\delta f_\alpha/{\bar{\tau}_{\alpha\beta}} - \delta{f_\beta}/{{\tau}_{\beta\alpha}}$. Note that the scattering times $\tau_{ij}$ and $\bar{\tau}_{ij}$ are symmetric when two antiferromagnetic magnon modes are degenerate, e.g., in the absence of a magnetic field. We solve the linearized Boltzmann equation for $\delta f_{\xi}$ in the steady-state limit and approximate linearly on the gradients of chemical potentials, where we considered the magnon chemical potential satisfies $\mu^e_{\alpha}+\mu^e_{\beta}=0$ \cite{Troncoso2020} at equilibrium and that the temperature is assumed homogeneous for simplicity. The magnon current, defined by ${\bs j}_{\xi}=(2\pi)^{-3}\hbar\int{d^3{\bs k}}\,\delta f_{\xi}{\bs v}_{\xi\bs k}$ with ${\bs v}_{\xi\bs k}=\partial\epsilon_{\alpha{\bs k}}/\partial{(\hbar\bs k)}$ the group velocity for the $\xi$-mode, is computed within the linear response theory and given by
\begin{align}\label{eq:jalphabeta}
 \left(\begin{array}{c}
         {\bs j}_{\alpha} \\
         {\bs j}_{\beta} 
         \end{array}\right)=
         \left(\begin{array}{cc}
         \sigma_{\alpha\alpha}     &  -\sigma_{\alpha\beta}\\
         -\sigma_{\beta\alpha}   & \sigma_{\beta\beta} 
         \end{array}\right)\left(\begin{array}{c}
         -\nabla\mu_{\alpha} \\
         -\nabla\mu_{\beta}
 \end{array}\right),
 \end{align}
where the elements of the magnon conductivity tensor are given by
\begin{subequations}
\begin{align}
\sigma_{\xi\xi}&\label{eq:sigma1}=\frac{\hbar\bar{\tau}_{\xi\xi}}{1-\Delta}\int\frac{d^3{\bs k}}{(2\pi)^3}\left(-\frac{\partial f^0_{\xi}}{\partial {\epsilon_{\xi\bs k}}}\right)|{\bs v}_{\bs k}|^2,\\
\sigma_{\xi\xi'}&\label{eq:sigma2}=\frac{\hbar\bar\tau_{\xi\xi'}\Delta}{1-\Delta}\int\frac{d^3{\bs k}}{(2\pi)^3}\left(-\frac{\partial f^0_{\xi'}}{\partial {\epsilon_{\xi'\bs k}}}\right)|{\bs v}_{\bs k}|^2,
\end{align}    
\end{subequations}
where $\Delta=\left[{\left(1+{\bar{\tau}_{\alpha\beta}}/{\bar{\tau}_{\alpha\alpha}}\right)\left(1+{\bar{\tau}_{\beta\alpha}}/{\bar{\tau}_{\beta\beta}}\right)}\right]^{-1}$, with $\bar{\tau}^{-1}_{\alpha\alpha}={\tau}^{-1}_{\alpha\alpha}+{\tau}^{-1}_{\alpha\beta}$ and $\bar{\tau}^{-1}_{\beta\beta}={\tau}^{-1}_{\beta\beta}+{\tau}^{-1}_{\beta\alpha}$. The magnon conductivity, $\sigma_{\xi\xi'}$, represents a drag effect between $\alpha$ and $\beta$ magnons, induced by interband magnon-magnon interactions \cite{Arakawa2022,Arakawa2022b}.  In the absence of the magnon interconversion, i.e., $\tau_{\xi\xi'}(\bar{\tau}_{\xi\xi'}) \to \infty$, the off-diagonal intermode conductivity $\sigma_{\xi\xi'}$ vanishes and the diagonal intramode term approaches the conductivity of purely diffusive magnons $\sigma^0_{\xi }\equiv\sigma_{\xi\xi}(\Delta=0)$ \cite{RezendePRB2016b,RezendePRB2016,Troncoso2020}.

\section{Hydrodynamic equations}
During collision events, conserved quantities satisfy a set of hydrodynamic equations derived from the corresponding conservation laws. These equations are obtained by averaging the conserved quantities over momentum space $\langle {\cal Q_{\xi}}\rangle_{\bs k}({\bs x},t)=\int d^3{\bs k}\,{\cal Q_{\xi}} f_{\xi}({\bs x},{\bs k},t)/\int d^3{\bs k}\,f_{\xi}({\bs x},{\bs k},t)$, where ${\cal Q_{\xi}}$ denotes a conserved quantity such as spin, linear momentum, or energy, and $f_{\xi}$ is the local equilibrium distribution function \cite{Chapman,Forster2018}. The relevant quantities for the $\xi$-magnons are magnon density, $\rho_{\xi}({\bs r},t)= \hbar V\int d^3{\bs k}/(2\pi)^3f_{\xi}({\bs x},{\bs k},t)$, and  the magnon drift velocity {${\bs v}_{\xi}({\bs r},t)=V\int d^3{\bs k}/(2\pi)^3\,(\hbar {\bs k}/(2m))\,f_{\xi}({\bs x},{\bs k},t)$, where $V$ is the system volume}. The hydrodynamic equations governing the nonequilibrium dynamics of  $\xi$-magnons, diffusion and Navier-Stokes equations, can be derived from the Boltzmann transport equation, Eq. (\ref{eq:BEalphabeta}).

The linearized magnonic diffusion equation, describing the spatial and temporal evolution of the magnon densities, takes the form of
\begin{align}\label{eq:denmag}
\dot{\rho}_{\xi}+\nabla\cdot\left(\rho_{\xi}{\bs v_{\xi}}\right)=-\frac{\hbar}{2e}\frac{\sigma^0_{\xi}}{\ell^2_{\xi}}\mu_{\xi}-g\left(\mu_{\xi}+\mu_{\xi'}\right),
\end{align}
with $\ell_{\xi}$ the magnon spin diffusion length \cite{Rezende2019,Cornelissen2016}, and $g$ describes the inelastic spin-conserving processes that account for, e.g., magnon-magnon and magnon-phonon scatterings, that keep the net magnon spin density $\sim \rho_{\alpha}-\rho_{\beta}$ constant, but the total number of magnons may change \cite{Troncoso2020}.  Secondly, the Navier-Stokes equation for the momentum densities, 
\begin{align}\label{eq:velmag}
\rho_{\xi}\left[\frac{\partial{\bs v}_{\xi}}{\partial t}+\left({\bs v}_{\xi}\cdot\nabla\right){\bs v}_{\xi}\right]\nonumber=&-\frac{\hbar}{2e}\frac{\sigma^0_{\xi}}{\tau_{\xi\xi}}\nabla\mu_{\xi}-\frac{\rho_{\xi}}{\tau_{\xi\xi}}{\bs v}_{\xi}-\frac{\rho_{\xi'}}{\tau_{\xi' \xi}}{\bs v}_{\xi'}\\
+&\eta_{\xi}\nabla^2{\bs v}_{\xi}+\eta_{\xi}'\nabla\left(\nabla\cdot{\bs v}_{\xi}\right),
\end{align}
where $\eta'_{\xi}=\chi_{\xi}+\eta_{\xi}/3$, with $\eta_{\xi}$ the dynamical shear viscosity and $\chi_{\xi}$ the bulk viscosity, derives from the momentum-conserving magnon-magnon interactions \cite{Batchelor2000}. The bulk viscosity, related to the resistance of the fluid to time-dependent volume changes, leads to a small renormalization of the magnon spin diffusion length and the magnon spin conductivity in the stationary regime. 

In the linear response and steady-state regime, the dynamics of the viscous magnon fluid are governed by a set of linearized diffusion and Navier–Stokes equations,
\begin{align}
\label{eq:diffusion}\nabla\cdot{\bs j}_{\xi}&=-\frac{\hbar\sigma^0_{\xi}}{2e\ell^2_{\xi}}\mu_{\xi}-g\left(\mu_{\xi}+\mu_{\xi'}\right),\\
\label{eq:navierstokes}{\bs j}_{\xi}&=-\frac{\hbar\sigma_{\nu\xi}}{2e}\nabla\mu_{\xi}-{{\chi}_{\xi'\xi}}\,{\bs j}_{\xi'}+{\cal D}^2_{\xi}\left(\nabla^2{\bs j}_{\xi}-\frac{g}{3}\nabla\mu_{\xi'}\right),
\end{align}
in terms of the linearized magnon current ${\bs j}_{\xi}=\rho_{\xi}{\bs v}_{\xi}$, where $\rho_{\xi}$ is the average magnon density obtained from the equilibrium Bose-Einstein distribution. We define ${{\chi}_{\xi'\xi}}={{\tau}_{\xi\xi}}/{{\tau}_{\xi'\xi}}$ and the prefactor ${\cal D}_{\xi}=\sqrt{\bar{\tau}_{\xi}{\nu_{\xi}}}$ as the momentum-relaxation length with $\nu_{\xi}={\eta_{\xi}}/{\rho_{\xi}}$ being the kinematic viscosity. In this regime, magnon transport exhibits both viscous and diffusive behavior. The magnon spin conductivity is given by $\sigma_{\nu\xi }/\sigma^0_{\xi}=1+\big({2e}g({\sigma^0_{\xi}\hbar})^{-1}+{ \ell^{-2}_{\xi}}\big){\cal D}^2_{\xi}/3$, modified by the factor $g$ and viscosity. The relevance of each effect derives from their characteristic length scales, i.e., the diffusion length $\ell_{\xi}$ and ${\cal D}_{\xi}$, which are determined by the viscosity and momentum relaxation. The magnon vorticity, $\bm{\omega}_{\xi}=\nabla\times{\bs j}_{\xi}$, measures the local fluid rotation, satisfies a steady-state diffusion-relaxation equation ${\cal D}^2_{\xi}\nabla^2\bm{\omega}_{\xi}-{{\chi}_{\xi'\xi}}\,\bm{\omega}_{\xi'}=\bm{\omega}_{\xi}$,  which is determined by the diffusion length and derived from Eqs. (\ref{eq:diffusion}) and (\ref{eq:navierstokes}). In the absence of viscosity, we recover the standard diffusive regime for the chemical potentials $\mu_{\xi}$ of antiferromagnetic magnons, 
\begin{align}\label{eq:effAFmagnondiff}
\nabla^2\mu_{\xi}-\frac{\sigma_{\xi\xi'}}{\sigma_{\xi\xi}}\nabla^2\mu_{\xi'}=\frac{\mu_{\xi}}{\Lambda^{2}_{\xi}}+\frac{\mu_{\xi'}}{\lambda^{2}_{\xi}},
\end{align}
with the magnon diffusion length ${\Lambda^{-2}_{\xi}}={\sigma^0_{\xi}}\sigma^{-1}_{\xi\xi}\ell^{-2}_{\xi}+{\lambda^{-2}_{\xi}}$, and ${\lambda^{-2}_{\xi}}={g}{\sigma^{-1}_{\xi\xi}}$. Note that Eq. (\ref{eq:effAFmagnondiff}), obtained by setting ${\cal D}_{\xi}=0$ and Eqs. (\ref{eq:diffusion}) and (\ref{eq:navierstokes}), differs from Ref. \cite{Troncoso2020} due to processes of interconversion of magnons quantified by $\sigma_{\xi\xi'}$. 

\begin{figure}[tbh]	\includegraphics[width=\columnwidth]{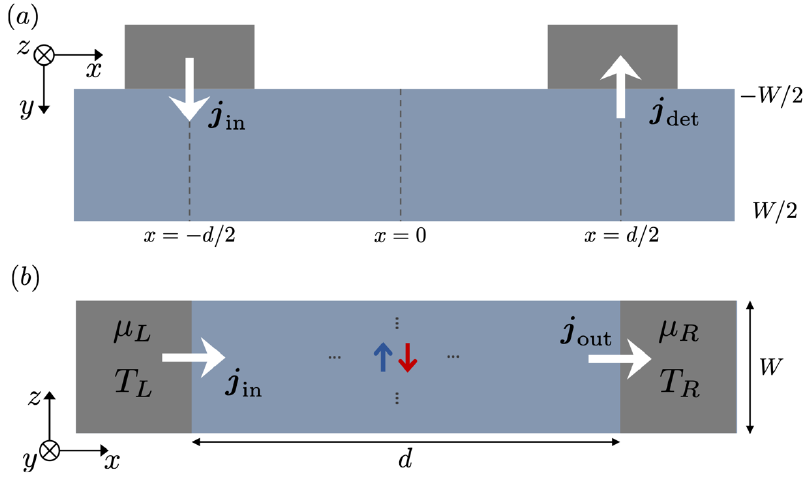}
	\caption{Schematic setups for  the (a) nonlocal transverse magnon transport and (b) longitudinal magnon transport through the NM-AF-NM heterostructure. The N\'eel ground state is along ${\bs z}$ direction, as displayed at panel (b). The identical left and right metallic leads are separated at distance $d$. Boundary conditions imposed at the interface with the metallic leads considers a difference between the magnon chemical potentials and a spin accumulation, induced by the SHE \cite{Jungwirth2016,Baltz2018rmp}.}
	\label{fig: magnon-transport}
\end{figure}

\section{Hydrodynamic Transport of Viscous Magnons in different setups}
Having established the governing equations for chiral magnon hydrodynamics in collinear AFs, we now numerically solve Eqs. (\ref{eq:diffusion}) and (\ref{eq:navierstokes}) with appropriate boundary conditions for two distinct geometries, as illustrated in Fig. \ref{fig: magnon-transport}. In panel (a), we illustrate a nonlocal \emph{transverse} magnon transport configuration, featuring two nonmagnetic metal leads positioned on the surface of the AF. In panel (b) a \emph{longitudinal} transport configuration is shown, comprising a metal–AF–metal heterostructure.

At finite temperature, an AF hosts a thermal magnon gas. When a spin voltage is applied via the SHE in a heavy-metal lead, spin is injected into the AF, generating a nonequilibrium population of $\xi$-magnons. {We consider the spin accumulation, ${\bs \mu}_s\cdot{\bs t}\mid_{\text{L}}\,=\mu_{\uparrow}-\mu_{\downarrow}$, induced at the left (L) lead, with $\mu_{\uparrow(\downarrow)}$ the spin-dependent chemical potential of electrons and ${\bs t}$ a unit vector pointing along the $\hat{\bs x}-$ and $\hat{\bs z}-$ direction, at setups (a) and (b) respectively.} {The injected magnon current across the interface is given in linear response by ${\bs j}_{\xi}\cdot{\bs n}\mid_{\text{L}}=g_s\left(\mu_s-\mu_{\xi}\right)$, with $g_s$ the interfacial spin conductance and the unit vector ${\bs n}$ normal to the boundary of the AF system.} The excited magnon mode $\xi$  is determined by the polarization direction of the spin voltage. These chiral magnons diffuse through the antiferromagnetic medium and are detected at the right (R) side by a second heavy-metal lead via the inverse SHE (ISHE) enabling the measurement of magnon transport. {For the right lead it is assumed that ${\bs \mu}_s\cdot{\bs t}\mid_{\text{R}}\,=0$, i.e., it behaves as an ideal spin sink}.

\subsection{Nonlocal transverse magnon transport}
The bulk continuity equations, Eqs. (\ref{eq:diffusion}) and (\ref{eq:navierstokes}), are complemented by a set of boundary conditions applied on the normal and tangential components of the spin current at the AF edges in the proposed geometry at Fig. \ref{fig: magnon-transport}(a). Firstly, at the nonlocal device, the magnon current is nonzero at the interface with the injector and detector, which are approximated as two $\delta$-functions localized at $x=\pm d/2$. The boundary condition at the upper surface is ${j}^{y}_{\xi}(x,-W/2)=g_s(\mu_s-\mu_{\xi}(x))\delta(x+d/2)+g_s\mu_{\xi}(x)\delta(x-d/2)$. The remaining interfaces are assumed opaque, thus the normal component of the current vanishes, i.e., ${j}^{y}_{\xi}(x,W/2)={j}^{x}_{\xi}(\pm L/2,y)=0$.
Additionally, for the tangential component of the magnon current, we use a slip boundary condition $j^x_{\xi}(x,\pm W/2) = \mp l_b\left[\partial_yj^x_{\xi}+\partial_xj^y_{\xi}\right]_{y=\pm W/2}$ and $j^y_{\xi}(\pm L/2,y)=\mp l_b\left[\partial_yj^x_{\xi}+\partial_xj^y_{\xi}\right]_{x=\pm L/2}$, which is characterized by a phenomenological boundary slip length $l_b$ \cite{TorrePRB2015} that quantifies a friction in the momentum of magnons moving parallel to the boundary. The limit $l_b\rightarrow\infty$ can be thought of as a frictionless magnon current slipping along the boundary. On the contrary, if $l_b\rightarrow 0$, the magnon current flow tangential to the edges vanishes at the boundary, which leads to a no-slip condition.
\begin{figure}[tbh]	
\includegraphics[width=\columnwidth]{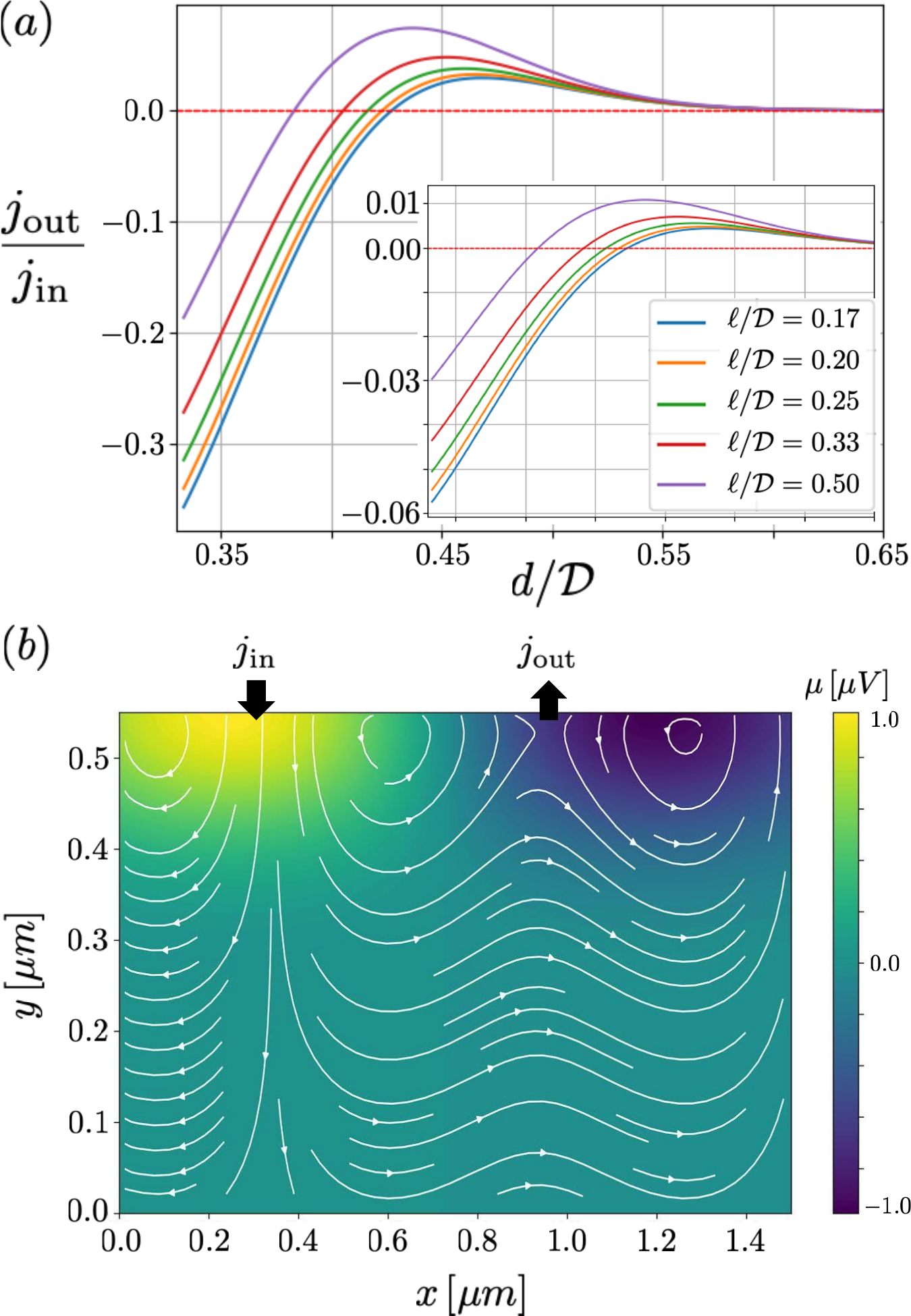}
\caption{(a) Nonlocal magnon spin current ratio $j_{\text{out}}/j_{\text{in}}$ as a function of the injector–detector separation, calculated in zero magnetic field for different magnon spin-diffusion lengths. The sample width is $W=0.1{\cal D}$ and the inelastic spin-conserving parameter is $g=2\times 10^{13}$ S/m$^{2}$. The inset shows the magnon current ratio for $g=0$. Each curve is plotted for different values of the slip length in the range $l_b\in[0.051\mu\text{m},0.15\mu\text{m}]$, from top to bottom. (b) Visualization of magnon current streamlines, with both incoming $j_{\text{in}}$ and outgoing $j_{\text{out}}$ spin currents, for the system with the following parameters: $l_b=0.1\mu\text{m}$ $\sigma^0_{\alpha}=10^{5}$ S/m, $\ell=0.1\mu$m, $g=g_s=10^{13}$ S/m$^{2}$, ${\cal D}=0.3\mu$m, $\tau^{-1}_{\alpha\alpha}=0.125$ Hz, $\tau^{-1}_{\alpha\beta}=0.25$ Hz.
The magnon chemical potential is depicted as the background of the image.}
\label{fig:nonlocal-current}
\end{figure}

The signature of a viscous magnon dynamics is determined by the ratio between the detected ($j_{\text{out}}$) and injected ($j_{\text{in}}$) magnon currents. This ratio is proportional to $R_{nl}/R_{0}$ \cite{Cornelissen2015}, where $R_{nl}$ and $R_{0}$ are the nonlocal resistance that is measured experimentally and the resistance of the leads, respectively. 

First, we assume a zero magnetic field and thus, the bulk magnon transport becomes identical for each mode, e.g., $\ell\equiv\ell_{\alpha}=\ell_{\beta}$ and ${\cal D}\equiv{\cal D}_{\alpha}={\cal D}_{\beta}$, since two magnon modes are degenerate. In Fig. \ref{fig:nonlocal-current}(a), we plot $j_{\text{out}}/j_{\text{in}}$ as a function of dimensionless distance between injector and detector, $d/{\cal D}$, and for several values of the magnon spin diffusion length. The results are obtained by numerical integration of Eqs. (\ref{eq:diffusion}) and (\ref{eq:navierstokes}) in the $xy$-plane, applying the corresponding boundary conditions and assuming translational invariance along the $z$ direction. A change in the sign of $j_{\text{out}}/j_{\text{in}}$ is observed in the presence of viscosity when the distance $d/{\cal D}$ decreases, similar to the ferromagnetic case \cite{UlloaPRL2019}. Interestingly, we show an enhancement (see inset of Fig. \ref{fig:nonlocal-current}(a) for $g=0$) of the ratio $j_{\text{out}}/j_{\text{in}}$ by the factor $g$, having a similar behavior to the viscosity; however, the maximum distance at which viscous effects are observed does not alter significantly. 

In Fig. \ref{fig:nonlocal-current}(b), the magnon chemical potential (depicted in color code) and a solution for the current streamlines are shown for $d/{\cal D}=0.35$ and $\ell/{\cal D}=0.33$.  We observe the magnon chemical potential becomes negative as the injector-detector distance is decreased. This behavior occurs near the region of the detector, when the ratio $j_{\text{out}}/j_{\text{in}}$ turns negative as shown at Fig. \ref{fig:nonlocal-current}(a), which is a consequence of the viscous effects. Additionally, the magnon current streamlines show a steady-state vortex flow, at the upper edge of the system near the detector. These vortices indicate a dynamic interaction between incoming and outgoing currents, causing variations in current density, where the current flowing upward towards the detector is marked by local circulations that generate differences in current density. The flow pattern reflects the complex interplay of system parameters and boundary effects. Note that in the absence of viscosity, the change of sign disappears and the standard diffusive regime is recovered \cite{Cornelissen2015}.
\begin{figure}[tbh]	\includegraphics[width=\columnwidth]{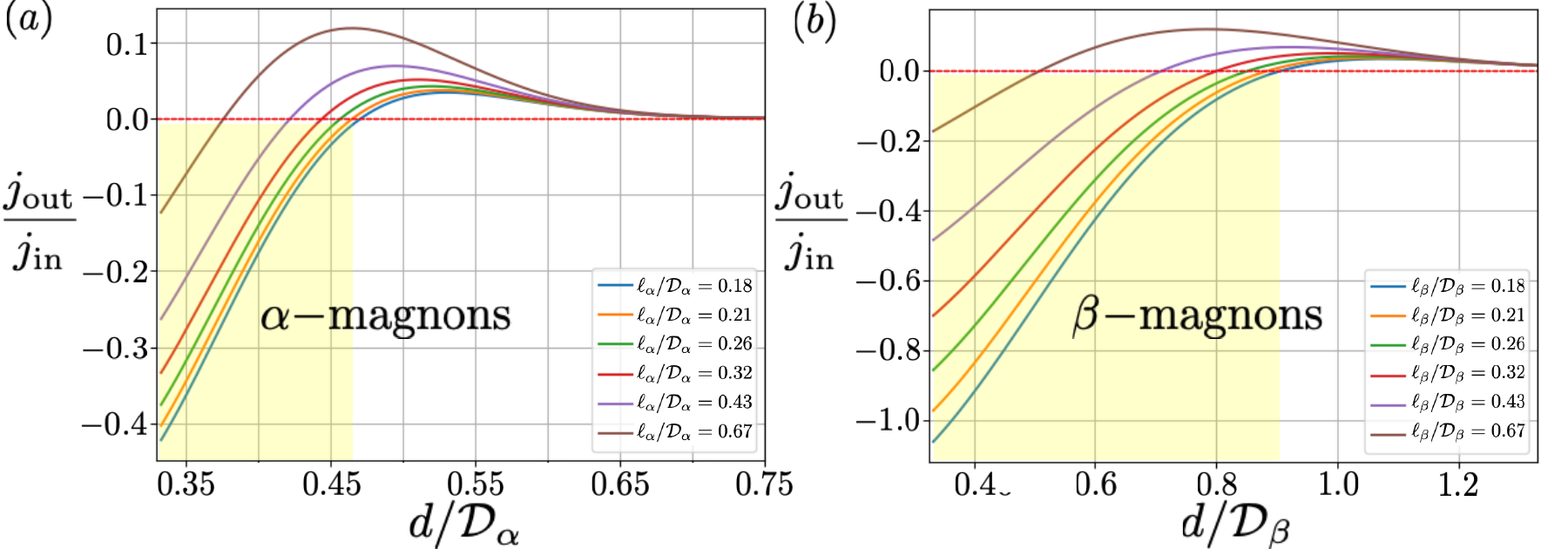}
	\caption{Magnon spin current ratio $j_{\text{out}}/j_{\text{in}}$ in the nonlocal setup as a function of distance in the presence of magnetic field for (a) $\alpha-$ and (b) $\beta-$magnons for different magnon spin-diffusion lengths.  The yellow area denote regions where $\alpha$-magnons (panel (a)) and $\beta$-magnons (panel (b)) exhibit a negative spin current ratio, indicative of the viscous regime. The curves at panel (a) and (b) are plotted for different values of the slip length in the range $l_b\in[0.018\mu\text{m},0.067\mu\text{m}]$, from top to bottom, and ${\cal D}_{\alpha}={\cal D}_{\beta}=0.1\mu$m.}	\label{fig4}
\end{figure} 

In the presence of a magnetic field along the magnetic ground state, the magnon currents associated with the $\alpha$ and $\beta$ modes become inequivalent, as the degeneracy between the two magnon modes is lifted. The effect on the magnon spin transport is displayed by the ratio $j_{\text{out}}/j_{\text{in}}$ of each magnon mode at Fig. \ref{fig4}. {The transport coefficients for each magnon mode become different in the presence of a magnetic field.} Our results show that at certain distances, the sign of the magnon spin-current ratio $j_{\mathrm{out}} / j_{\mathrm{in}}$ differs between the $\alpha$- and $\beta$-magnon modes, revealing distinct transport regimes for each mode. When $\alpha$-magnons display a positive ratio, indicative of diffusive transport, $\beta$-magnons show a negative ratio, characteristic of the viscous regime, and vice versa. The regions where each mode exhibits viscous behavior are highlighted by the yellow areas in Fig. \ref{fig4}(a) and (b). Reversing the magnetic-field direction switches the transport character of the $\alpha$-magnons, demonstrating that both the direction and magnitude of the field control the viscous–diffusive crossover independently for each magnon mode.

\subsection{Longitudinal magnon transport: Magnonic Poiseuille flow} 
We now consider the longitudinal transport setup, consisting of a NM$|$AF$|$NM trilayer heterostructure, as illustrated in Fig. \ref{fig: magnon-transport}(b), in the absence of a magnetic field. In this geometry, the magnon currents satisfy the boundary conditions ${j}_{\xi}^{x}(x=-W/2)=G_\xi\left[\mu_{L}\mp\mu_\xi(-W/2)\right]$ and ${j}_{\xi}^{x}(x=W/2)=G_\xi\left[\mu_{R}\mp\mu_\xi(W/2)\right]$. The spin accumulation at the interface of the normal metals and AF is denoted by $\mu_L$ and $\mu_R$, whose interfaces are located at $x=-W/2$ and $x=W/2$, respectively. The contact magnon conductances are $G_{\alpha}$ and $G_{\beta}$ at each interface \cite{Troncoso2020}. Furthermore, the magnon currents perpendicular to the upper and lower interfaces vanish, ${j}_{\xi}^{z}(x,z=0)={j}_{\xi}^{z}(x,z=W)=0$, and similar slip boundary conditions apply along the edges of this geometry. As in the previous case, we ignore the temperature gradient and thus the spin Seebeck effect.
\begin{figure}[tbh]
\includegraphics[width=1\columnwidth]{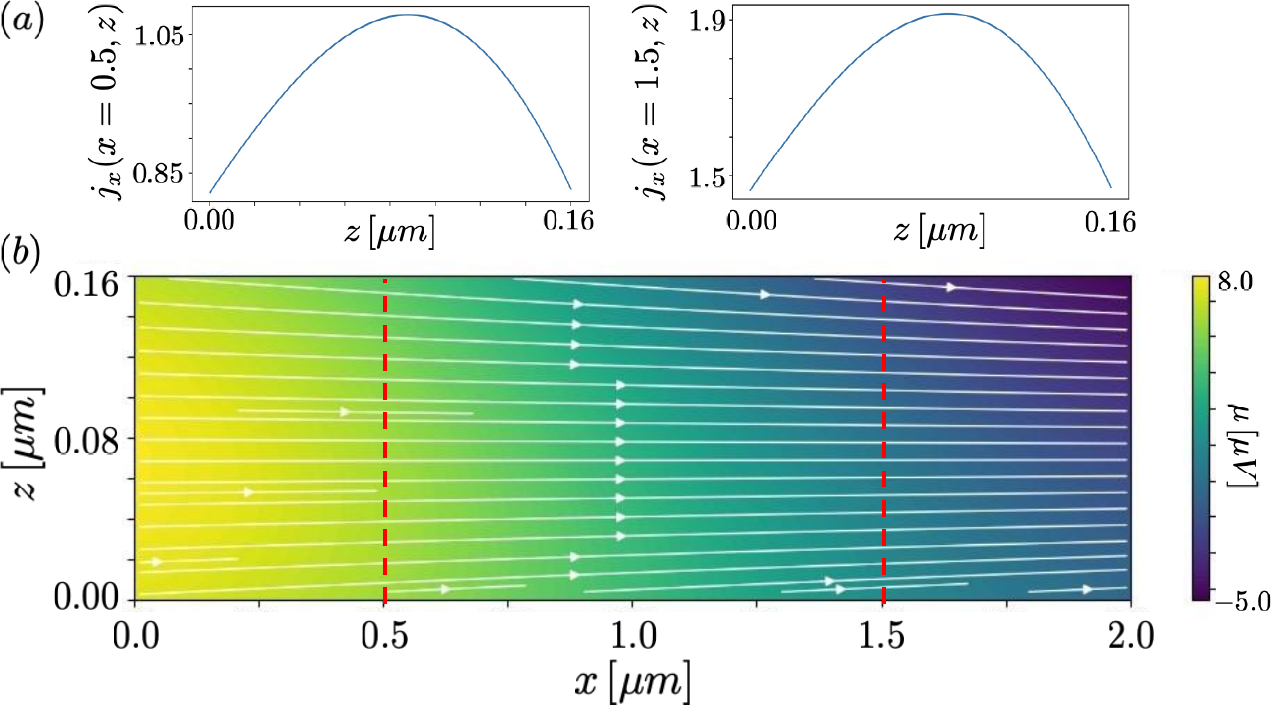}
\caption{Current streamlines in the longitudinal setup, demonstrating the characteristic parabolic profile of Poiseuille flow, shown at panel (b). The white streamlines illustrate the velocity profile, with their concavity (shown on panel (a) at different positions) providing insight into to flow dynamics. In regions with high velocity gradients, the streamlines exhibit inward curvature. In contrast, regions with lower velocity gradients show outward concavity, indicating flow expansion as the fluid moves away from the boundaries. The set of parameters used are, $l_b=0.2\mu$m $\sigma_{\alpha 0}=10^{5}$ S/m, $\ell=0.2\mu$m, $G_{\alpha}=G_{\beta}=10^{13}$ S/m$^{2}$, ${\cal D}=0.3\mu$m, $\tau^{-1}_{\alpha\alpha}=0.125$ Hz, and $\tau^{-1}_{\alpha\beta}=0.25$ Hz.
The magnon chemical potential is depicted as the background of the image.}
	\label{fig:poiseuille flow}
\end{figure}

The observation of magnonic Poiseuille flow, suggests that magnons behave as a viscous fluid rather than diffusing incoherently, see Fig. \ref{fig:poiseuille flow}, which requires dominant momentum-conserving magnon-magnon interactions over momentum-relaxing processes like impurity or boundary scattering. Poiseuille flow implies a parabolic spatial profile of magnon current across the sample width, shown at Fig. \ref{fig:poiseuille flow}(a),  analogous to the flow of a classical viscous fluid through a pipe \cite{Batchelor2000}. This behavior is a hallmark of collective magnon dynamics and indicates that magnons can transport spin in a coherent and long-range manner. 
In the Poiseuille regime, spin conductivity can be significantly enhanced compared to diffusive regimes, offering a route to low-dissipation spin transport. This can lead to the design of efficient magnonic devices that exploit collective behavior rather than individual magnon motion. Detecting Poiseuille flow involves identifying nonlocal transport signals and spatial profiles that deviate from standard diffusive expectations. Techniques such as nonlocal spin Seebeck measurements or imaging of spin current profiles are valuable in this context. 
\\
\section{Conclusion and Discussion}
In this paper, we investigated the dynamics of chiral magnons in a collinear AF within the hydrodynamic regime in two different setups. Particularly, we focused on the transport of spin angular momentum and linear momentum, which are described by diffusion and Navier-Stokes equations for the density of chiral magnons and their respective drift velocities. This regime is established when the momentum-relaxation length is larger than the momentum-conserving scattering processes. 

Signatures from the hydrodynamic regime lead to changes in magnon transport measurements, such as nonlocal resistance measurements and in spin-conductance transport experiments. In particular, we found that collisions between chiral magnons carrying opposite spin-angular momentum affect the propagation of spin currents in the form of drag-type effects. A vortex pattern in the magnon spin currents is found in the nonlocal geometry, while a Poiseuille flow of magnons is observed only in the longitudinal setup. Our results allow probing interaction-dominated magnon transport, for which the magnon mean free path due to collisions is smaller than the device dimension or impurity mean free path.

The emergence of viscous flow in chiral magnon systems necessitates ultraclean magnetic materials, cryogenic temperatures, and mesoscopic sample sizes, where momentum-conserving scattering dominates over momentum-relaxing mechanisms. Theoretically, this phenomenon bridges magnonics and fluid dynamics, paving the way for interdisciplinary exploration. Technologically, it offers the prospect of robust, scalable spin transport systems resilient to conventional charge-scattering effects. 

\section{Acknowledgments}
R.E.T. and V. O. A. thanks funding from Fondecyt Regular 1230747. A. Q. thanks funding from the Research Council of Norway through its Centers of Excellence funding scheme, Project No. 262633 ``QuSpin'' and FRIPRO with Project No. 353919 ``QTransMag''.

\bibliography{hydromagnonAF}

\onecolumngrid

\appendix
\section{Magnon-magnon interactions}\label{app:scattamp}
The spin Hamiltonian at Eq. (\ref{eq:spinH}) is expanded up to fourth order in the magnon operators via HP transformation \cite{HolsteinPR1940}. Then, Fourier transformed through the relations $a_{\bs r_i}=\frac{1}{\sqrt{N}}\sum_{i}e^{i{\bs k}\cdot{\bs r_i}}a_{\bs k}$ and $b_{\bs r_j}=\frac{1}{\sqrt{N}}\sum_{j}e^{i{\bs k}\cdot{\bs r_j}}b_{\bs k}$
we find the interacting magnon Hamiltonian  ${H}_{AF}=E_0+H_{0}+H_{I}$, where
\begin{align}
H_{0}&\label{eq:H2AF}=(Jsz+\kappa)\sum_{\bf q}\left[(1+h)a^{\dagger}_{\bf q}a_{\bf q}+(1-h)b^{\dagger}_{\bf q}b_{\bf q}+\xi\gamma_{\bf q}(a_{\bf q}b_{-\bf q}+a^{\dagger}_{\bf q}b^{\dagger}_{-\bf q})\right]\\
H_{I}&=\nonumber-\frac{Jz}{2N}\sum_{{\bf q}_1{\bf q}_2{\bf q}_3{\bf q}_4}\delta_{{\bf q}_1+{\bf q}_2-{\bf q}_3-{\bf q}_4}\left[2\gamma_{{\bf q}_2-{\bf q}_4}a^{\dagger}_{\bf q_1}b^{\dagger}_{-\bf q_4}a_{\bf q_3}b_{-\bf q_2}+\frac{\kappa}{2s}\left(a^{\dagger}_{\bf q_1}a^{\dagger}_{\bf q_2}a_{\bf q_3}a_{\bf q_4}+b^{\dagger}_{\bf q_1}b^{\dagger}_{\bf q_2}b_{\bf q_3}b_{\bf q_4}\right)\right.\\
&\label{eq:H4AF}\left.\qquad\qquad\qquad\qquad\qquad\qquad+\gamma_{{\bf q}_4}\left(b^{\dagger}_{{\bf q}_1}b_{-{\bf q}_2}b_{{\bf q}_3}a_{{\bf q}_4}+b^{\dagger}_{{\bf q}_3}b^{\dagger}_{-{\bf q}_2}b_{{\bf q}_1}a^{\dagger}_{{\bf q}_4}+a^{\dagger}_{\bf q_1}a_{-\bf q_2}a_{\bf q_3}b_{{\bf q_4}}+a^{\dagger}_{\bf q_3}a^{\dagger}_{-\bf q_2}a_{\bf q_1}b^{\dagger}_{{\bf q_4}}\right)\right]
\end{align}
with $h=H/(Jsz+\kappa)$, $\xi=Jsz/(Jsz+\kappa)$ and $\gamma_{\bf q}=\frac{2}{z}\sum_{{\bs \delta}}\cos\left[{\bf q}\cdot{\bs \delta}\right]$ where $z$ is the coordination number. The quadratic part of the Hamiltonian, Eq. (\ref{eq:H2AF}), is diagonalized by the Bogoliubov transformation
\begin{align}
\hat{a}_{\bs q}&\label{eq: bogoliubov1}=l_{\bs q}\hat{\alpha}_{\bs q}+m_{\bs q}\hat{\beta}^{\dagger}_{-\bs q}\\
\hat{b}^{\dagger}_{-\bs q}&\label{eq: bogoliubov2}=m_{\bs q}\hat{\alpha}_{\bs q}+l_{\bs q}\hat{\beta}^{\dagger}_{-\bs q}
\end{align}
with the coefficients $l_{\bf q}=\left(\frac{(Jsz+\kappa)+\epsilon_{\bf q}}{2\epsilon_{\bf q}}\right)^{1/2}$, $m_{\bf q}=-\left(\frac{(Jsz+\kappa)-\epsilon_{\bf q}}{2\epsilon_{\bf q}}\right)^{1/2}\equiv-\chi_{\bf q}l_{\bf q}$ and $\epsilon_{\bf q}=(Jsz+\kappa)\sqrt{1-\xi^2\gamma^2_{\bf q}}$, resulting in Eq. (\ref{eq:freemagnonH}). In the diagonal basis, the interacting Hamiltonian becomes 
\begin{align}\label{eq:magmaghamiltonian}
H_I=\sum_{\mathbf{q}_1{\bf q}_2{\bf q}_3{\bf q_4}}&\nonumber\delta_{{\bf q}_1+{\bf q}_2-{\bf q}_3-{\bf q}_4}\left[V^{(1)}_{{\bf q}_1{\bf q}_2{\bf q}_3{\bf q}_4}\alpha^{\dagger}_{{\bf q}_1}\alpha^{\dagger}_{{\bf q}_2}\alpha_{{\bf q}_3}\alpha_{{\bf q}_4}+V^{(2)}_{{\bf q}_1{\bf q}_2{\bf q}_3{\bf q}_4}\alpha^{\dagger}_{{\bf q}_1}\beta_{-{\bf q}_2}\alpha_{{\bf q}_3}\alpha_{{\bf q}_4}+V^{(3)}_{{\bf q}_1{\bf q}_2{\bf q}_3{\bf q}_4}\alpha^{\dagger}_{{\bf q}_1}\alpha^{\dagger}_{{\bf q}_2}\alpha_{{\bf q}_3}\beta^{\dagger}_{-{\bf q}_4}\right.\\
&\nonumber\left.+V^{(4)}_{{\bf q}_1{\bf q}_2{\bf q}_3{\bf q}_4}\alpha^{\dagger}_{{\bf q}_1}\beta_{-{\bf q}_2}\alpha_{{\bf q}_3}\beta^{\dagger}_{-{\bf q}_4}+V^{(5)}_{{\bf q}_1{\bf q}_2{\bf q}_3{\bf q}_4}\beta_{-{\bf q}_1}\beta_{-{\bf q}_2}\alpha_{{\bf q}_3}\beta^{\dagger}_{-{\bf q}_4}+V^{(6)}_{{\bf q}_1{\bf q}_2{\bf q}_3{\bf q}_4}\alpha^{\dagger}_{{\bf q}_1}\beta_{-{\bf q}_2}\beta^{\dagger}_{-{\bf q}_3}\beta^{\dagger}_{-{\bf q}_4}\right.\\
&\left.+V^{(7)}_{{\bf q}_1{\bf q}_2{\bf q}_3{\bf q}_4}\alpha^{\dagger}_{{\bf q}_1}\alpha^{\dagger}_{{\bf q}_2}\beta^{\dagger}_{-{\bf q}_3}\beta^{\dagger}_{-{\bf q}_4}+V^{(8)}_{{\bf q}_1{\bf q}_2{\bf q}_3{\bf q}_4}\beta_{-{\bf q}_1}\beta_{-{\bf q}_2}\alpha_{{\bf q}_3}\alpha_{{\bf q}_4}+V^{(9)}_{{\bf q}_1{\bf q}_2{\bf q}_3{\bf q}_4}\beta_{-{\bf q}_1}\beta_{-{\bf q}_2}\beta^{\dagger}_{-{\bf q}_3}\beta^{\dagger}_{-{\bf q}_4}\right],
\end{align}
where the scattering amplitudes are $V^{(a)}_{{\bf q}_1{\bf q}_2{\bf q}_3{\bf q}_4}=-\left(\frac{Jz}{N}\right)l_{{\bf q}_1}l_{{\bf q}_2}l_{{\bf q}_3}l_{{\bf q}_4}\Phi^{(a)}_{{\bs 1}{\bs 2}{\bs 3}{\bs 4}}$. The functions $\Phi^{(a)}$ are the following expressions

\begin{align}
\Phi^{(1)}_{{\bs 1}{\bs 2}{\bs 3}{\bs 4}}&\label{eq:phi1}=\gamma_{{\bf q}_2-{\bf q}_4}\chi_{{\bf q}_2}\chi_{{\bf q}_4}-\frac{1}{2}\left(\gamma_{{\bf q}_2}\chi_{{\bf q}_2}+\gamma_{{\bf q}_4}\chi_{{\bf q}_4}+\gamma_{{\bf q}_2}\chi_{{\bf q}_1}\chi_{{{\bf q}_3}}\chi_{{{\bf q}_4}}+\gamma_{{\bf q}_4}\chi_{{{\bf q}_1}}\chi_{{{\bf q}_2}}\chi_{{{\bf q}_3}}\right)+\frac{\kappa}{2Jzs}\left(1+\chi_{{\bf q}_1}\chi_{{\bf q}_2}\chi_{{\bf q}_3}\chi_{{\bf q}_4}\right)
\\
\Phi^{(2)}_{{\bs 1}{\bs 2}{\bs 3}{\bs 4}}&\label{eq:phi2}\nonumber=-\gamma_{{\bf q}_2-{\bf q}_4}\chi_{{\bf q}_4}-\gamma_{{\bf q}_1-{\bf q}_4}\chi_{{\bf q}_1}\chi_{{\bf q}_2}\chi_{{\bf q}_4}+\gamma_{{\bf q}_4}\chi_{{{\bf q}_1}}\chi_{{{\bf q}_3}}+\gamma_{{\bf q}_4}\chi_{{{\bf q}_2}}\chi_{{{\bf q}_4}}+\frac{1}{2}\left(\chi_{{{\bf q}_3}}\chi_{{{\bf q}_4}}\left(\gamma_{{\bf q}_1}+\gamma_{{\bf q}_2}\chi_{{\bf q}_1}\chi_{{\bf q}_2}\right)+\left(\gamma_{{\bf q}_2}+\gamma_{{\bf q}_1}\chi_{{{\bf q}_1}}\chi_{{{\bf q}_2}}\right)\right)\\
&-\frac{\kappa}{Jzs}\left(\chi_{{\bf q}_2}+\chi_{{\bf q}_1}\chi_{{\bf q}_3}\chi_{{\bf q}_4}\right)
\\
\nonumber\Phi^{(3)}_{{\bs 1}{\bs 2}{\bs 3}{\bs 4}}&\label{eq:phi3}=-\gamma_{{\bf q}_2-{\bf q}_4}\chi_{{\bf q}_2}-\gamma_{{\bf q}_2-{\bf q}_3}\chi_{{\bf q}_2}\chi_{{\bf q}_3}\chi_{{\bf q}_4}+\gamma_{{\bf q}_2}\chi_{{\bf q}_1}\chi_{{\bf q}_3}+\gamma_{{\bf q}_2}\chi_{{\bf q}_2}\chi_{{\bf q}_4}+\frac{1}{2}\left(\chi_{{\bf q}_1}\chi_{{\bf q}_2}\left(\gamma_{{\bf q}_3}+\gamma_{{\bf q}_4}\chi_{{\bf q}_3}\chi_{{\bf q}_4}\right)+\left(\gamma_{{\bf q}_4}+\gamma_{{\bf q}_3}\chi_{{\bf q}_3}\chi_{{\bf q}_4}\right) \right)\\
&-\frac{\kappa}{Jzs}\left(\chi_{{\bf q}_4}+\chi_{{\bf q}_1}\chi_{{\bf q}_2}\chi_{{\bf q}_3}\right)
\\
\Phi^{(4)}_{{\bs 1}{\bs 2}{\bs 3}{\bs 4}}&\label{eq:phi4}\nonumber=\gamma_{{\bf q}_2-{\bf q}_4}+\gamma_{{\bf q}_1-{\bf q}_4}\chi_{{\bf q}_1}\chi_{{\bf q}_2}+\gamma_{{\bf q}_2-{\bf q}_3}\chi_{{\bf q}_3}\chi_{{\bf q}_4}+\gamma_{{\bf q}_1-{\bf q}_3}\chi_{{\bf q}_1}\chi_{{\bf q}_2}\chi_{{\bf q}_3}\chi_{{\bf q}_4}+\frac{2\kappa}{Jzs}\left(\chi_{{\bf q}_2}\chi_{{\bf q}_4}+\chi_{{\bf q}_1}\chi_{{\bf q}_3}\right)\\
&-\left(\chi_{{\bf q}_1}\left(\gamma_{{\bf q}_3}+\gamma_{{\bf q}_4}\chi_{{\bf q}_3}\chi_{{\bf q}_4}\right)+\chi_{{\bf q}_3}\left(\gamma_{{\bf q}_1}+\gamma_{{\bf q}_2}\chi_{{\bf q}_1}\chi_{{\bf q}_4}\right) +\chi_{{\bf q}_4}\left(\gamma_{{\bf q}_2}+\gamma_{{\bf q}_1}\chi_{{\bf q}_1}\chi_{{\bf q}_2}\right) +\chi_{{\bf q}_2}\left(\gamma_{{\bf q}_4}+\gamma_{{\bf q}_3}\chi_{{\bf q}_3}\chi_{{\bf q}_4}\right)\right)
\\
\Phi^{(5)}_{{\bs 1}{\bs 2}{\bs 3}{\bs 4}}\nonumber&\label{eq:phi5}=-\gamma_{{\bf q}_2-{\bf q}_4}\chi_{{\bf q}_1}-\gamma_{{\bf q}_2-{\bf q}_3}\chi_{{\bf q}_1}\chi_{{\bf q}_3}\chi_{{\bf q}_4}+\gamma_{{\bf q}_2}\chi_{{\bf q}_2}\chi_{{\bf q}_3} +\gamma_{{\bf q}_2}\chi_{{\bf q}_1}\chi_{{\bf q}_4}+\frac{1}{2}\left(\left(\gamma_{{\bf q}_3}+\gamma_{{\bf q}_4}\chi_{{\bf q}_3}\chi_{{\bf q}_4}\right)+\chi_{{\bf q}_1}\chi_{{\bf q}_2}\left(\gamma_{{\bf q}_4}+\gamma_{{\bf q}_3}\chi_{{\bf q}_3}\chi_{{\bf q}_4}\right)\right)\\
&-\frac{\kappa}{Jzs}\left(\chi_{{\bf q}_3}+\chi_{{\bf q}_1}\chi_{{\bf q}_2}\chi_{{\bf q}_4}\right)
\\
\Phi^{(6)}_{{\bs 1}{\bs 2}{\bs 3}{\bs 4}}&\label{eq:phi6}\nonumber=-\gamma_{{\bf q}_2-{\bf q}_4}\chi_{{\bf q}_3}-\gamma_{{\bf q}_1-{\bf q}_4}\chi_{{\bf q}_1}\chi_{{\bf q}_2}\chi_{{\bf q}_3}+\gamma_{{\bf q}_4}\chi_{{\bf q}_1}\chi_{{\bf q}_4}+ \gamma_{{\bf q}_4}\chi_{{\bf q}_2}\chi_{{\bf q}_3}+\frac{1}{2}\left(\left(\gamma_{{\bf q}_1}+\gamma_{{\bf q}_2}\chi_{{\bf q}_1}\chi_{{\bf q}_2}\right)+\chi_{{\bf q}_3}\chi_{{\bf q}_4}\left(\gamma_{{\bf q}_2}+\gamma_{{\bf q}_1}\chi_{{\bf q}_1}\chi_{{\bf q}_2}\right)\right)\\
&-\frac{\kappa}{Jzs}\left(\chi_{{\bf q}_1}+\chi_{{\bf q}_2}\chi_{{\bf q}_3}\chi_{{\bf q}_4}\right)
\\
\Phi^{(7)}_{{\bs 1}{\bs 2}{\bs 3}{\bs 4}}&\label{eq:phi7}=\gamma_{{\bf q}_2-{\bf q}_4}\chi_{{\bf q}_2}\chi_{{\bf q}_3}-\frac{1}{2}\left(\gamma_{{\bf q}_2}\chi_{{\bf q}_1}+\gamma_{{\bf q}_4}\chi_{{{\bf q}_3}}+\gamma_{{\bf q}_4}\chi_{{{\bf q}_1}}\chi_{{{\bf q}_2}}\chi_{{{\bf q}_4}}+\gamma_{{\bf q}_2}\chi_{{{\bf q}_2}}\chi_{{{\bf q}_3}}\chi_{{{\bf q}_4}}\right)+\frac{\kappa}{2Jzs}\left(\chi_{{\bf q}_3}\chi_{{\bf q}_4}+\chi_{{\bf q}_1}\chi_{{\bf q}_2}\right)
\\
\Phi^{(8)}_{{\bs 1}{\bs 2}{\bs 3}{\bs 4}}&\label{eq:phi8}=\gamma_{{\bf q}_2-{\bf q}_4}\chi_{{\bf q}_1}\chi_{{\bf q}_4}-\frac{1}{2}\left(\gamma_{{\bf q}_4}\chi_{{{\bf q}_3}}+\gamma_{{\bf q}_2}\chi_{{{\bf q}_1}}+\gamma_{{\bf q}_2}\chi_{{{\bf q}_2}}\chi_{{{\bf q}_3}}\chi_{{{\bf q}_4}}+\gamma_{{\bf q}_4}\chi_{{{\bf q}_1}}\chi_{{{\bf q}_2}}\chi_{{{\bf q}_4}}\right)+\frac{\kappa}{2Jzs}\left(\chi_{{\bf q}_1}\chi_{{\bf q}_2}+\chi_{{\bf q}_3}\chi_{{\bf q}_4}\right)
\\
\Phi^{(9)}_{{\bs 1}{\bs 2}{\bs 3}{\bs 4}}&\label{eq:phi9}=\gamma_{{\bf q}_2-{\bf q}_4}\chi_{{\bf q}_1}\chi_{{\bf q}_3}-\frac{1}{2}\left(\gamma_{{\bf q}_4}\chi_{{{\bf q}_4}}+\gamma_{{\bf q}_2}\chi_{{{\bf q}_2}}+\gamma_{{\bf q}_2}\chi_{{{\bf q}_1}}\chi_{{{\bf q}_3}}\chi_{{{\bf q}_4}}+\gamma_{{\bf q}_4}\chi_{{{\bf q}_1}}\chi_{{{\bf q}_2}}\chi_{{{\bf q}_3}}\right)+\frac{\kappa}{2Jzs}\left(1+\chi_{{\bf q}_1}\chi_{{\bf q}_2}\chi_{{\bf q}_3}\chi_{{\bf q}_4}\right)
\end{align}
where $\chi_{{\bf q}}=-\left(\frac{1-\epsilon_{\bf q}}{1+\epsilon_{\bf q}}\right)^{1/2}$. Note the symmetry relations among these coefficients $\Phi^{(3)}_{{\bs 1}{\bs 2}{\bs 3}{\bs 4}}=\Phi^{(2)}_{{\bs 3}{\bs 4}{\bs 1}{\bs 2}}$, $\Phi^{(6)}_{{\bs 1}{\bs 2}{\bs 3}{\bs 4}}=\Phi^{(5)}_{{\bs 3}{\bs 4}{\bs 1}{\bs 2}}$ and $\Phi^{(8)}_{{\bs 1}{\bs 2}{\bs 3}{\bs 4}}=\Phi^{(7)}_{{\bs 3}{\bs 4}{\bs 1}{\bs 2}}$.  Thus, the scattering amplitudes at Eq. (\ref{eq:interactingV1}) and (\ref{eq:interactingV2}), are given by
\begin{align}
{\textsc v}^{(\alpha)}_{{\bs q}{\bs k}{\bs k}'}&=-\left(\frac{Jz}{N}\right)l_{{\bs k}+{\bs q}}l_{{\bs k}'-{\bs q}}l_{{\bs k}}l_{{\bs k}'}\Phi^{(1)}_{{\bs k}+{\bs q},{\bs k}'-{\bs q},{\bs k},{\bs k}'},\\
{\textsc v}^{(\beta)}_{{\bs q}{\bs k}{\bs k}'}&=-\left(\frac{Jz}{N}\right)l_{{\bs k}}l_{{\bs k}'}l_{{\bs k}+{\bs q}}l_{{\bs k}'-{\bs q}}\Phi^{(9)}_{{\bs k},{\bs k}',{\bs k}+{\bs q},{\bs k}'-{\bs q}},\\
u_{{\bs q}{\bs k}{\bs k}'}&=-\left(\frac{Jz}{N}\right)l_{{\bs k}+{\bs q}}l_{{\bs k}'-{\bs q}}l_{{\bs k}}l_{{\bs k}'}\Phi^{(2)}_{{\bs k}+{\bs q},{\bs k}'-{\bs q},{\bs k},{\bs k}'},\\
v_{{\bs q}{\bs k}{\bs k}'}&=-\left(\frac{Jz}{N}\right)l_{{\bs k}+{\bs q}}l_{-{\bs k}'-{\bs q}}l_{{\bs k}}l_{-{\bs k}'}\Phi^{(4)}_{{\bs k}+{\bs q},-{\bs k}'-{\bs q},{\bs k},-{\bs k}'},\\
w_{{\bs q}{\bs k}{\bs k}'}&=-\left(\frac{Jz}{N}\right)l_{{\bs q}-{\bs k}'}l_{{\bs k}-{\bs q}}l_{{\bs k}}l_{-{\bs k}'}\Phi^{(5)}_{{\bs q}-{\bs k}',{\bs k}-{\bs q},{\bs k},-{\bs k}'}.
\end{align}

\end{document}